\documentstyle[12pt]{article}

\newcommand{\be}{\begin{equation}}
\newcommand{\ee}{\end{equation}}
\newcommand{\bea}{\begin{eqnarray}}
\newcommand{\eea}{\end{eqnarray}}

\begin{document}
\begin{titlepage}


\vspace{1in}

\begin{center}
\Large
{\bf T-DUALITY OF NSR SUPERSTRING: THE WORLDSHEET PERSPECTIVE }

\vspace{1in}

\normalsize

{ Jnanadeva Maharana \footnote{Adjunct Professor, National Institute of 
Science Education and Research, Bhubaneswar, India } \\ 
E-mail maharana$@$iopb.res.in 
\\}
 
\noindent \today

\normalsize
\vspace{.5in}

 {\em Institute of Physics \\
Bhubaneswar - 751005 \\
India  \\ }

\end{center}

\vspace{1in}

\baselineskip=24pt
\begin{abstract}
\noindent We formulate  target space duality symmetry of NSR superstring
from the perspectives of worldsheet.  
The worldsheet action is presented in the superspace formalism 
in the presence of massless 
backgrounds. We start from a ${\hat D}$-dimensional target space worldsheet
action and compactify the theory on a d-dimensional torus, $T^d$.  
It is assumed
that the backgrounds are independent of compact (super)coordinates. We
adopt the formalism of our earlier work to introduce dual supercoordinates
along compact directions and introduce 
the corresponding dual backgrounds. It is demonstrated
that  combined equations of motion of the two sets of coordinates can be
expressed in a manifestly $O(d,d)$ covariant form analogous to the equations
of motions for closed bosonic string derived by us. Furthermore, we show
that the vertex operators associated with excited massive levels of NSR
string can be expressed in an $O(d,d)$ invariant form generalizing earlier
result for closed bosonic string.

\end{abstract}

\vspace{.7in}

\end{titlepage}



\noindent{\bf 1. INTRODUCTION}

\bigskip

\noindent
The rich symmetry contents of string theory have played a cardinal role
in understanding diverse attributes of string theory and have
provided deep incisive insights into its dynamics. The target space duality
(T-duality) is a very special feature of string theory and has attracted
considerable attentions over two decades \cite{books,rev}. 
This symmetry owes its existence
to one-dimensional nature of string and we do not encounter the analog of
T-duality in field theoretic description of a point particle. Therefore,
it is natural to explore the symmetry from the perspective of the worldsheet.
The T-duality symmetry, associated with closed bosonic string, has been 
investigated from the worldsheet perspective in a general framework almost
two decades ago \cite{ms}. The closed bosonic string was 
considered in the presence of
its massless backgrounds and $d$ of its target space coordinates were
compactified on a torus. The backgrounds along compact directions were allowed
to depend on noncompact string coordinates and were assumed 
to be independent of compact coordinates. It was recognized that the worldsheet
equations of motion are conserved currents along compact directions. We
introduced dual coordinates and corresponding dual backgrounds to derive
another set of equations of motion. The two sets of equations of motion were
be suitably combined to derived $O(d,d)$ covariant equations \cite{ms}
where $d$ is the
number of compact coordinates.\\ 
We may recall that the first hint of the T-duality was unraveled in the
context of Regge-phenomenology in the context of strong interactions in the
garb of FESR duality. Therefore, intuitively, it is quite tempting to
conjecture that the
excited stringy levels might possess some of the attributes of T-duality. It
might be useful to explore T-duality symmetry of massive states from the
perspective of the worldsheet. Indeed,
existence of such a symmetry for excited massive levels of closed
bosonic string \cite{maha} was demonstrated in the sense that the vertex
operators for these levels were cast in an $O(d,d)$ invariant form. These 
results were derived in in a simple scenario .\\
On the other hand when T-duality is analyzed in a general setting its
salient features and powerful applications are exhibited from the view point
of the target space in the effective action approach. 
To recapitulate, when we
envisage evolution of a closed string in the background of its massless
excitations and demand (quantum) conformal invariance the backgrounds are
constrained through the $\beta$-function equations which are computed
perturbatively in the worldsheet $\sigma$-model approach. These equations of
motion enable us to introduce the effective action whose variation reproduces
the "equations of motion". Let us  toroidally compactify  the
effective action to lower dimensions and examine symmetries of the reduced
effective action. The reduced action can be cast in a manifestly $O(d,d)$
invariant form following the Scherk-Schwarz \cite{ss} 
dimensional reduction scheme
if the theory is compactified on a d-dimensional torus ($T^d$) and the
massless backgrounds are independent of the compact coordinates whereas they
carry spacetime dependence. 
The availability of a manifestly $O(d,d)$ invariant reduced effective action
has been very useful to explore various aspects of string theory in diverse
directions. The target space duality has attracted a lot of attentions
 from different  
perspectives over a long period. We refer to some early papers \cite{revodd}
and 
interested reader may consult reviews for comprehensive list of 
papers \cite{rev}.\\
The purpose of this investigation is primarily two fold: We investigate
target space duality symmetry of NSR superstring from the perspective of the
worldsheet. A comprehensive understanding of $O(d,d)$ symmetry for NSR string
in this framework is lacking. We adopt the worldsheet superspace approach
to study the T-duality. We demonstrate that an $O(d,d)$ covariant  worldsheet
evolution 
equation can be obtained for NSR superstring. In order to achieve our 
objectives we express the NSR action in the superspace 
(in superconformal gauge) in the presence of its massless excitations such as
graviton and two-form antisymmetric tensor. When the target space is 
compactified on $T^d$ and the backgrounds are independent of these 
super coordinates, we generalize our earlier results \cite{ms} 
for closed string, to
cast the corresponding worldsheet equations of motion in a manifestly
$O(d,d)$ covariant form. \\
The other goal is to examine existence of T-duality
symmetry for excited massive levels of NSR string when the theory is 
compactified on $T^d$. We achieve this objective
 by introducing T-duality doublet
of vectors in superspace which are the basic building blocks of all vertex
operators; thus substantially improving earlier results on T-duality for 
massive states \cite{maha}.
We introduce a set of projection operators to
express vertex operators of massive levels in an $O(d,d)$ invariant form.
As an illustrative example we consider compactification of type IIB theory to
$AdS_3 \otimes S^3 \otimes T^4$ and focus our attention to NS-NS sector. We
show, in this case, how the T-duality invariant vertex operators can be
constructed for the excited massive levels of superstring.\\
Let us very briefly recapitulate some of the essential results of T-duality 
from the worldsheet point of view in the case of closed bosonic string. To
begin with, consider the worlsheet action in the presence of constant 
backgrounds, ${\hat G}_{\hat\mu\hat\nu}$ and $ {\hat B}_{\hat\mu\hat\nu}$
\bea
\label{freeaction}
S={1\over 2}\int d\sigma d\tau \bigg({\gamma^{ab}}
{\hat G}_{{\hat \mu}{\hat{\nu}}}\partial_a X^{\hat{\mu}}\partial_b
X^{\hat{\nu}}+\epsilon^{ab}{\hat B}_{{\hat \mu}{\hat{\nu}}}
\partial_a X^{\hat{\mu}}\partial_b X^{\hat{\nu}}    \bigg)
\eea
where $\gamma^{ab}={\rm diag}(1,-1)$ worldsheet metric in the orthonormal
gauge, ${\hat\mu},{\hat\nu}=0,1,2,... {\hat D}-1$,
 ${\hat G}_{{\hat\mu}{\hat\nu}}$ is 
a constant  metric of the target space, ${\hat B}_{{\hat \mu}{\hat{\nu}}}$
is the constant antisymmetric tensor; consequently, the presence of the last 
term in
(\ref{freeaction}) does not contribute to the equations of motion.
The duality symmetry is manifestly exhibited  if we express the resulting
canonical Hamiltonian density in the form
\be
\label{hamiltonian}
{\hat H}_c= {1\over 2}{\hat V}^T{\hat M}{\hat V}
\ee
where
\bea
\label{oddvector}
{\hat V} =\pmatrix{P_{\hat \mu} \cr X'^{\hat \mu} \cr}
\eea
$P_{\hat \mu}$ being the conjugate momenta and prime denoting the 
$\sigma$-derivative. Moreover, the $2{\hat D}\times2 {\hat D}$ symmetric
matrix \cite{mmatrix,mmatrix2}
\bea
\label{matrix}
{\hat M} = \pmatrix {{\hat G}^{-1} & -{\hat G}^{-1}{\hat B} \cr
{\hat B}{\hat G}^{-1} & {\hat G} - {\hat B}{\hat G}^{-1} {\hat B}\cr} \eea
is defined in terms of the constant backgrounds appearing in the worldsheet
action above. The Hamiltonian density 
(\ref{hamiltonian}) is invariant under the 
global $O({\hat D}, {\hat D})$ transformation
\bea
\label{odd2}
{\hat M}\rightarrow {\hat\Omega}{\hat M}\Omega ^T,~~
{\hat V}\rightarrow {\hat\Omega} {\hat V},~~ 
{\hat\Omega} ^T{\hat\eta}{\hat\Omega} ={\hat\eta},~~
{\hat\Omega} \in O({\hat D},{\hat D}),~ 
{\hat\eta}=\pmatrix{0 & {\bf 1} \cr {\bf 1} & 0 \cr}
\eea
$\bf \hat \eta$ is the $O({\hat D},{\hat D})$ metric and $\bf 1$ is 
${\hat D}\times {\hat D}$ unit matrix
 and ${\hat V} $ is the $O({\hat D},{\hat D})$ vector and 
${\hat M}\in O({\hat D},{\hat D})$. This symmetry is generalization of the 
duality symmetry $P_{\hat\mu}\leftrightarrow X'^{\hat\mu}$. 
In the framework of Lagrangian
formulation, the $O({\hat D}, {\hat D})$ symmetry is exhibited as follows.
In presence of constant backgrounds, the woldsheet equations of motion for
string coordinates $\{ X^{\hat \mu} ({\sigma, \tau}) \}$ are set 
of conservation
laws
\be
\partial _a{\cal J}^a_{\hat\mu}=0
\ee
where the current is given by 
\bea
{\cal J}^a_{\hat\mu}=\gamma^{ab}{\hat G}_{{\hat\mu}{\hat\nu}}
\partial_b X^{\hat\nu}+
\epsilon^{ab}{\hat B}_{{\hat\mu}{\hat\nu}}\partial_b X^{\hat\nu} 
\eea 
Thus
locally, one can express the two dimensional current as: 
\be
{\cal J}^a_{\hat\mu}=\epsilon_{ab}\partial^b{\tilde Y}_{\hat{\mu}} 
\ee
where
$\{{\tilde Y}_{\hat\mu} \}$ are set of dual coordinates. The next step is
to introduce a set of dual backgrounds ${\widetilde G}$ and 
${\widetilde B}$ and 
finally a dual action. The equations of motion for ${\tilde Y}_{\hat\mu}$
involving dual backgrounds is also a conservation law. The two sets of
conservation laws (equations of motion consisting of $2{\hat D}$ vectors) can 
be combined to derive a single  $O({\hat D}, {\hat D})$ covariant equations
of motion \cite{duff}.
 Subsequently, this result has been derived for the cosmological
case i.e. when the backgrounds $G$ and $B$ are time dependent \cite{mahat}.\\
In the more general case, alluded to in the introduction,
 where the the massless
backgrounds assume spacetime dependence the following results hold.
Let us compactify the string on a d-dimensional torus $T^d$ and decompose
the worldsheet string coordinates into two sets 
\be
X^{\hat\mu}=(X^{\mu}, Y^{\alpha}),~~\mu=0,1,..D-1;~\alpha=1,2,..d 
\ee
so that ${\hat D}=D+d$. $X^{\mu}$ and $Y^{\alpha}$ are the spacetime
and compact coordinates respectively. We suppress explicit dependence of these
coordinates on $\sigma$ and $\tau$ from now on.
For such a toroidal compactification we can
decompose the backgrounds as
\bea
\label{schwarz}
{\hat e}^{\hat r}_{\hat\mu}=\pmatrix{ e^r_{\mu}(X) & 
A^{(1)\beta}_{\mu}(X)E^a(X)_{\beta}(X)\cr 0 & E^a_{\alpha}(X)\cr }
\eea
The spacetime metric is $g_{\mu\nu}=e^r_{\mu}g^{(0)}_{rs}e^s_{\nu}$ and
the internal metric is $G_{\alpha\beta}=E^a_{\alpha}\delta_{ab}E^b_{\beta}$;
$g^{(0)}_{rs}$ is the D-dimensional flat space Lorentzian signature metric.
$ A^{(1)\beta}_{\mu}$ are gauge fields associated with the d isometries and
it is assumes that the backgrounds depend of coordinates $X^{\mu}$ and are
independent of $Y^{\alpha}$. Similarly, the antisymmetric tensor background,
depending only on $X^{\mu}$ can be decomposed as
\bea 
{\hat B}_{{\hat\mu}{\hat\nu}} =\pmatrix{B_{\mu\nu}(X) & B_{\mu\alpha}(X)\cr
B_{\nu\beta}(X) & B_{\alpha\beta}(X) \cr} 
\eea 
Here we note the presence of gauge fields $ B_{\mu\alpha}$ due to 
compactification as expected.
\\
We go through the steps of deriving equations of motion for the compact
coordinates, $Y^{\alpha}$ from the $\sigma$-model action and note that these
still correspond to conservation laws since backgrounds and gauge fields are
independent of compact string coordinates. Although the set of equations are
more complicated in the general setting, eventually, after long and
tedious calculations, the worldsheet equations of motion can be cast in 
an $O(d,d)$ covariant form as was demonstrated by us \cite{ms}.\\
There is another interesting approach to dualities in the worldsheet
approach. In this formulation the number of string coordinates are doubled and
this in scenario some of the nice features of conventional worldsheet approach 
are not maintained; however, it has been argued that such doubling might
have deep significance \cite{duff,witten} in string theory.
 Recently, a new formulation
of  field theory has been introduced where $O(D,D)$ invariant action is
constructed, $D$ being the number of spacetime dimensions which is doubled
\cite{double1,double2}. At this stage we have not been able to establish 
connection of our formulation with double field theory.

We mention
{\it en passant} that the case of superstring is not straight forward
if we deal with worldsheet bosonic coordinates and their fermionic partners.
We address this case in the next section.

\bigskip

\noindent {\bf 2. T-DUALITY OF COMPACTIFIED NSR STRING} 

\bigskip

\noindent A free superstring ( NSR string) action can be
expressed as sum of actions for set of left moving and right moving bosons
and fermions. Therefore, unlike the closed bosonic string case, we do not
see the $P\leftrightarrow X'$ duality (which is same as 
$\sigma \leftrightarrow\tau$ duality) so explicitly in the resulting
Hamiltonian density in the presence of fermionic coordinates. 
In fact the more transparent duality symmetry is to study 
the transformation properties of left moving and right moving fields under
$\sigma\leftrightarrow \tau$ interchange. The holomorphic fields do not
change sign whereas antiholomorphic ones do. If we introduce constant 
backgrounds as in the case of closed bosonic string the analog of noncompact 
$O({\hat D}, {\hat D})$ symmetry does not emerge so neatly. The target space
duality for superstrings in the NSR formulation has been studied in the past, 
however, we feel that this problem deserves further attention.
\\
Let us recapitulate evolution of NSR string in the background of massless
excitations. One starts with the superworldsheet action in two dimensional
superspace where the components of superfield are the bosonic coordinates,
(NSR) Majorana 
fermions and auxiliary fields and the backgrounds are functions of the
superfields. We expand the backgrounds in terms of component fields, eliminate
the auxiliary fields  in order to arrive at NSR superstring action in the
presence of massless backgrounds with component 
the fermionic and boson fields only. 
If we envisage the case where backgrounds
are independent of some of the coordinates (now backgrounds and their
derivatives depend only on bosonic coordinates), then it is very hard to 
arrive at duality invariant/covariant equations as was achieved by Maharana and
Schwarz \cite{ms}. Das and Maharana \cite{dm} considered NSR string action in 
superspace and adopted the technique introduced by Maharana and Schwarz 
\cite{ms} for the closed bosonic string to get analogous equations of motion.
However, they were unable to arrive at at duality covariant equations of 
motion although they obtained interesting results for a special case. 
In this case the $Z_2$ duality conditions are recovered
\be
\label{z2duality}
\partial_{\pm}X^{\mu}\rightarrow \pm\partial_{\pm}X^{\mu},
~~~\psi^{\mu}_{\pm}\rightarrow \pm\psi^{\mu}_{\pm}
\ee
Moreover, Siegel \cite{warren} considered superstring in superspace in a
Hamiltonian phase space approach to study dualities. Subsequently, there
have been attempts to reveal duality symmetries on superstring \cite{rest}.
Thus far worldsheet equations of motion for superstrings respecting target 
space duality symmetry has not been derived in a systematic manner at par
with the results of closed bosonic string.
\\
The NSR superstring action in two dimensional superspace is given by
\bea
\label{supaction}
S=-{1\over 2}\int d\sigma d\tau d^2\theta{\overline D}{\hat \Phi}^{\hat \mu}
\bigg({\hat G}_{{\hat\mu}{\hat\nu}}({\hat\Phi})-\gamma_5
{\hat B}_{{\hat\mu}{\hat\nu}}({\hat\Phi})\bigg)D{\hat\Phi}^{\hat{\nu}}
\eea
We have adopted superorthonormal gauge in arriving at this action. Here
${\hat G}_{\hat{\mu\nu}}({\hat \Phi}) $ and 
${\hat B}_{\hat{\mu\nu}}({\hat \Phi})$ are the graviton and and 2-form 
backgrounds which depend on the superfield ${\hat\Phi}$. It has 
expansion in component fields as
\bea
\label{superfield}        
{\hat \Phi}^{\hat\mu}= X^{\hat\mu}+{\bar\theta}\psi^{\hat\mu}+
{\bar\psi}^{\hat\mu}\theta+
{1\over 2}{\bar\theta}\theta  F^{\hat\mu}
\eea
where $ X^{\hat\mu}, \psi^{\hat\mu} ~{\rm and}~  F^{\hat\mu}$
are the bosonic, fermionic and auxiliary fields respectively. The covariant
derivatives in superspace are defined to be
\bea
\label{superder}
D_{\alpha}={{\partial}\over{\partial{\bar\theta_{\alpha}}}}-
i(\gamma^a\theta)_{\alpha}\partial_a,~~
{\overline D}_{\alpha}=-{{\partial}\over{\partial\theta_{\alpha}}}+
i({\bar\theta}\gamma^a)_{\alpha}\partial_a
\eea
where $\partial_a$ stands for worldsheet derivatives ($\sigma$ and $\tau$)
and the convention for $\gamma$ matrices are
\bea
\label{gamma}
\gamma^0=\pmatrix{0 & 1\cr 1 & 0\cr},~ 
\gamma^1=\pmatrix{0 & -1\cr 1 & 0\cr},~
\gamma_5=\gamma^0\gamma^1=\pmatrix{1 & 0\cr 0 & -1\cr}
\eea
The resulting equations of motion from (\ref{supaction}) are 
\bea
\label{eq}
{\overline D}\bigg({\hat G}_{{\hat\mu}{\hat\nu}}({\hat\Phi})-\gamma_5
{\hat G}_{{\hat\mu}{\hat\nu}}({\hat\Phi})\bigg)D{\hat\Phi}^{\hat\nu}=0
\eea
The equations for component fields can be obtained by expanding the backgrounds
in terms of them and utilizing the definitions of superspace derivatives
(\ref{superder}). Let us consider a compactification \cite{hs}
scheme such that target
space is compactified on $T^d$: ${\hat{\cal M}}_{\hat D}=M_D\otimes T^d$.
The metric and 2-form backgrounds are decomposed as
\bea
\label{scheme}
{\hat G}_{{\hat\mu}{\hat\nu}}=\pmatrix{g_{\mu\nu}(\phi) & 0\cr 0 & 
G_{ij}(\phi)\cr}, ~ 
{\hat B}_{{\hat\mu}{\hat\nu}}=\pmatrix{B_{\mu\nu}(\phi) & 0 \cr 
0 &  B_{ij}(\phi) \cr}
\eea
Note that the backgrounds only depend on spacetime superfields, $\phi^{\mu}$. 
We decompose the superfields as
\be
{\hat\Phi}^{\hat\mu}=(\phi^{\mu}, W^i)
\ee
where $\mu, \nu=0,1,2..D-1$ and $i,j=1,2,..d$ with ${\hat D}=D+d$. Note that
the two superfields have the expansions
\bea
\label{spacetimef1}
 \phi^{\mu}= X^{\mu}+{\bar\theta}{\psi}^{\mu}+
{\bar\psi}^{\mu}\theta+{1\over 2}{\bar\theta}\theta F^{\mu}
\eea
and
\bea
\label{compactf}
 W^i= Y^i+{\bar\theta}{\chi}^i+
{\bar\chi}^i\theta+{1\over 2}{\bar\theta}\theta F^i
\eea
$\chi^i$ are two dimensional Majorana spinors. 
In this compactification scheme, the equations of motion for the superfields
$\phi^{\mu}$ is exactly analogous to (\ref{eq}) where we replace $\hat\Phi$
with $\phi$ and the backgrounds with $g_{\mu\nu}(\phi)$ and $B_{\mu\nu}(\phi)$.
\\
Let us focus attention on the evolution equations and the 
dynamics of superfields along compact directions.
The action is 
\bea
\label{waction}
S=-{1\over 2}\int d\sigma d\tau d^2\theta{\overline D}W^i \bigg(G_{ij}(\phi)-
\gamma_5B_{ij}(\phi) \bigg)DW^j
\eea
The superderivatives are defines in (\ref{superder}) above. 
The equations of motion for $\{ W^i \}$ are 
\bea
\label{wieq}
{\overline D}\bigg(\bigg[G_{ij}(\phi)-\gamma_5 B_{ij}(\phi)\bigg]DW^j\bigg)=0
\eea
In view of the fact that $G$ and $B$ depend only on $\phi^{\mu}$, we may
introduce a dual free superfield ${\widetilde W}_i$ satisfying following 
 equation locally which is consistent with (\ref{wieq}) 
\bea
\label{dualsf}
\bigg(G_{ij}(\phi)-\gamma_5 B_{ij}(\phi) \bigg)DW^j=D{\widetilde W}_i
\eea
and the dual superfield satisfies the constraint: 
${\overline D}D{\widetilde W}_i=0$.
Note that (\ref{dualsf}) is reminiscent of the dual coordinate introduced
for closed string by us in the case of closed closed string under a similar
 scenario \cite{ms}. We can go further and
opt for a first order formalism and consider the Lagrangian density
\bea
{\widetilde{\cal L}}={1\over 2}{\bar\Sigma}^i\bigg(G_{ij}(\phi)-
\gamma_5B_{ij}(\phi)
\bigg)\Sigma ^j -{\bar\Sigma}^i D{\widetilde W}_i
\eea
The ${\bar\Sigma}^i$ variation leads to
\be
\bigg(G_{ij}(\phi)-\gamma_5 B_{ij}(\phi) \bigg)\Sigma ^j=D{\widetilde W}_i
\ee
and ${\widetilde W}_i$ variation implies ${\overline D}{\bar \Sigma}^i=0$. 
Therefore,
when $\Sigma ^i=DW^i$ we recover (\ref{wieq}). 
\\
We are in a position to introduce a dual Lagrangian density in terms of
the dual superfields, ${\widetilde W}_i$ and a set of 
dual backgrounds ${\cal G}^{ij}(\phi)$
and ${\cal B}^{ij}(\phi)$; whereas the former of the two backgrounds is 
symmetric in its indices the latter is antisymmetric.
\bea
\label{dualw}
{\cal L}_{\widetilde W}= -{1\over 2}{\overline D}{\widetilde W}_i
\bigg({\cal G}^{ij}(\phi)
-\gamma_5 {\cal B}^{ij}(\phi) \bigg)D{\widetilde W}_j
\eea
where the two dual backgrounds, $({\cal G}, {\cal B})$, are related to the
original background fields, $(G, B)$ through the following equations 
\bea
\label{dualbg}
{\cal G}=\bigg(G-BG^{-1}B \bigg)^{-1} ~~{\rm and}~~ 
{\cal B}=-\bigg(G-BG^{-1}B \bigg)^{-1}BG^{-1}
\eea
Notice that $(G-BG^{-1}B)^{-1}$ is symmetric since $(G-BG^{-1}B)$ is symmetric
and it is easy to check that ${\cal B}$ is antisymmetric.
The equations of motion associated with (\ref{dualw}) is
\bea
\label{wideq}
{\overline D}\bigg(\bigg[{\cal G}(\phi)-\gamma_5{\cal B}(\phi) \bigg]
D{\widetilde W}\bigg)=0
\eea
 Our next step is to
write down a pair of equations relating the superfields and their duals
which will lead us to T-duality covariant equations of motion. This is 
facilitated
by inspecting the two sets of equations of motion (\ref{wieq}) and 
(\ref{wideq})
 resulting from the original 
Lagrangian density and its dual which correspond to two   conservation laws. 
After straight forward and a little tedious calculations 
we arrive at following two
equations
\bea
\label{odd1}
DW^i=\gamma_5(G^{-1}B)^i_jDW^j+{G^{-1}}^{ij}D{\widetilde W}_j
\eea
\bea
\label{odd2}
D{\widetilde W}_i=\gamma_5({\cal G}^{-1}{\cal B})^j_iD{\widetilde W}_j+
{\cal G}^{-1}_{ij}DW^j
\eea 
Using (\ref{dualbg}) in (\ref{odd2}) we get two sets of equations relating
$DW^i$ and $D{\tilde W}_i$. Let us define a 2d-dimensional $O(d,d)$ vector
(each one is a superfield) such that
\be
\label{oddvec}
{\bf U}=\pmatrix{W^i\cr {\widetilde W}_i\cr} 
\ee
and a matrix
\bea
\label{msmatrix}
{\cal M}=\pmatrix{{\bf 1}G^{-1} & \gamma_5G^{-1}B \cr
-\gamma_5 BG^{-1} & {\bf 1}G-{\bf 1}BG^{-1}B \cr}
\eea
where ${\bf 1}$ is the $2\times 2$ unit matrix and $\gamma_5$ is two 
dimensional diagonal matrix defined earlier. The ${\cal M}$ matrix has
properties of the familiar $M$-matrix introduced in dimensional reduction of
closed bosonic string: ${\cal M}\in O(d,d)$ and corresponding metric is $\eta$.
The dimensions are further doubled due to the presence of two component 
Majorana
fermions.and is reflected by the appearance of $\bf 1$ and $\gamma_5$ in
the ${\cal M}$-matrix.   
The two equations (\ref{odd1}) and (\ref{odd2}) can be combined to 
a single matrix equation
\bea
\label{oddeqn}
D{\bf U}={\cal M}\eta{\bf U}
\eea
It follows from the definition of the $O(d,d)$ vector ${\bf U}$ that
${\overline D}D{\bf U}=0$. 
It holds by virtue of the fact that the two components of
${\bf U}$ satisfy  ${\overline D}DW^i=0$ and ${\overline D}D{\widetilde W}_i=0$
from our original equations (they are dual superfields of each other).
Therefore, we arrive at an $O(d,d)$ covariant equations of motion for 
coordinates along compact directions
\bea
\label{finalodd}
{\overline D}\bigg({\cal M}\eta{\bf U}\bigg)=0
\eea
Thus (\ref{finalodd}) generalizes the closed string $O(d,d)$ covariant 
equations of motion to NSR superstring.
\\
Let us return to the evolution equation for the superfields corresponding
to noncompact coordinates
\bea
\label{noncompact}
{\overline D}\bigg(g_{\mu\nu}(\phi)-
\gamma_5 B_{\mu\nu}(\phi) \bigg)D\phi_{\nu}=0
\eea
Notice that due to the dependence of backgrounds on the superfield $\phi$
these are "dynamical" equations unlike the case of compact coordinates which
were identified as conservation laws. More important point to note is
that  these equations are T-duality
invariant since the background tensors and these superfields are inert 
under the
action of the T-duality noncompact symmetry group. 
Therefore, we conclude that the
resulting equations of motion for a superstring compactified on $T^d$ can be
cast in an $O(d,d)$ covariant form. In the next section, we shall consider
an illustrative example.

\bigskip

\noindent {\bf 3. TYPE IIB COMPACTIFICATION ON}
 ${\bf AdS_3\otimes S^3\otimes T^4}$

\bigskip

\noindent
We envisage a scenario where our results can be concretely realized. In the
presence of NS-NS 3-form flux we can write down an worldsheet action for the
case at hand. Notice that $AdS_3$ and $S^3$ correspond to target space of
constant negative and positive curvatures respectively. Therefore, if we
introduce appropriate NS-NS three form fluxes, we can describe the Lagrangian
in these two sectors as sum of two WZW Lagrangians. The presence of WZ term
renders the theory conformally invariant and has the interpretation of the 
background antisymmetric tensor fields. Moreover, for the present scenario the
the associated field strengths are constant. The radii of these two spaces are
to be such that the cosmological constants arising from constant positive
and negative curvatures of $S^3$ and $AdS_3$ correspondingly sum up to zero. 
The worldsheet description of
NSR string on $AdS_3\otimes S^3$ can be formulated as WZW model on group
manifolds $SL(2,R)\otimes SU(2)$ as is well known. Thus the full worldsheet
action is
decomposed into sum of three parts: one corresponds to superconformal
theory on $SL(2,R)$ the other being $SU(2)$ and the third part is the one
describing a supersymmetric
$\sigma$-model along compact direction as we have discussed
in the previous section.\\
Let us briefly consider bosonic WZW model for $SU(2)$ group whose action is
\bea
\label{su2}
S_B &&={1\over {4\lambda^2}}\int d\sigma d\tau{\rm Tr}
\bigg(\partial_ag^{-1}(\sigma,\tau)\partial^ag(\sigma,\tau)\bigg)\nonumber\\
&&+ {k\over{16\pi}}\int_B{\rm Tr}\bigg(g^{-1}(\sigma,\tau)dg(\sigma,\tau)\wedge
g^{-1}(\sigma,\tau)dg(\sigma,\tau)\wedge g^{-1}(\sigma,\tau)dg(\sigma.\tau)
\bigg)
\eea
where $g\in SU(2)$ and satisfies the constraint $gg^{\dagger}={\bf 1}$, 
${\bf 1}$ being the unit matrix. 
Note the following features: (i) $\lambda$ and $k$ are dimensionless coupling
constants. For a compact group like $SU(2)$, $k$, the coupling constant 
appearing in front of the WZ term is quantized for the consistency of the 
quantized theory. (ii) g should be smoothly extended to a 3-dimensional 
manifold and its boundary, B, is the worldsheet (actually one should define
complex variables in terms of $(\sigma, \tau)$ and this action in those 
variables in the standard manner.) The theory is conformally invariant at the
special point ${\lambda ^2}={{4\pi}\over k}$. \\
The case of (bosonic) string on noncompact $SL(2,R)$ manifold is similar to
$SU(2)$ with some differences. (i) The matrix ${\tilde g}\in SL(2,R)$
satisfies the constraint ${\tilde g}\zeta{\tilde g}^T=\zeta$ to be contrasted
with $g\in SU(2)$ group element. Here $\zeta$ is the $SL(2,R)$ metric with 
property $\zeta ^2=-{\bf 1}$ and it can be chosen to be
\be
\zeta=\pmatrix{0 & 1\cr -1 & 0\cr}
\ee
(ii) In this case the coefficient of WZ, $k$ term need not be quantized.
\\
We shall consider the supersymmetric WZW model for $SU(2)$ from now on.
The action is \cite{a1,a2,a3,revwzw}
\bea
\label{superwzw}
S={1\over {4\lambda^2}}\int d\sigma d\tau d^2\theta{\bar D}{\bf G}^{\dagger}
D{\bf G} 
+{k\over{16\pi}}\int d\sigma d\tau d^2\theta\int^1_0dt
{\bf G}^{\dagger}{{{\bf dG}}\over{dt}} {\bar D}{\bf G}^{\dagger}
\gamma_5 D{\bf G}
\eea
The matrix ${\bf G}$ defined in the 
superspace satisfies constraints ${\bf G}{\bf G}^{\dagger}={\bf 1}$.
In order to define the WZ term as an integral over a three dimensional
space one defines extension of the superfield to 3-dimensions so that $t=0$
corresponds to the boundary i.e. at that point the two dimensional superfield
is defined on the worldsheet and two dimensional $\gamma_5$ is defined already.
Several remarks are in order at this point: (i)  When the 
${\bf G}\in SU(2)$ matrix is expressed in terms of component fields, auxiliary
field is eliminated, the $d\theta$ integration is done, the resulting action
contains quartic fermionic coupling and the theory is not necessarily
conformally invariant for arbitrary $\lambda^2$ and $k$. (ii) At the special
point $\lambda^2={{4\pi}\over k}$, theory is conformally invariant and the
quartic fermionic coupling disappears. Therefore, for a superstring on a group
manifold the two coupling constants are related (and $k$ is quantized). 
Moreover, at the conformal point, the equations of motion of the superfields
(${\bf G}$-matrices), decompose into holomorphic and antiholomorphic parts and
they take the form of two separate current conservation equations. This feature
is most elegantly displayed if we expand the super-matrix in ${\bf G}$ 
in light cone variables as
\bea
\label{lcexp}
{\bf G}(\sigma,\tau,\theta)
=g(\sigma,\tau)\bigg( {\bf 1}+i\theta^+\psi_+(\sigma,\tau)
+i\theta^-\psi_-(\sigma,\tau)+i\theta^+\theta^-F(\sigma,\theta) \bigg)
\eea
Corresponding light cone superderivatives are
\bea
\label{superlc}
D_{\pm}={{\partial}\over{\partial\theta^{\pm}}}-i\theta^{\pm}\partial_{\pm}
~~{\rm with}~~\partial_{\pm}=\partial_{\tau}\pm\partial_{\sigma}
\eea
Note that the chiral fermions $\psi_{\pm}$ are matrices taking value in the 
Lie algebra and $F$ is the auxiliary field. The constraint $GG^{\dagger}=
{\bf 1}$ results in relations between component fields, $g,\psi_{\pm}$ and
$F$. At the conformal point the equation of motions become
\bea
\label{seqn}
D_{\mp}{\cal {\bf J}}_{\pm}=0, ~~~~
{\cal{\bf J }}_{\pm}=-i{\bf G}^{-1}D_{\pm}{\bf G}
\eea
We therefore, note that classically we solve for NSR string on $S^3$.\\
The case of NSR string on $AdS_3$, proceeds similarly once we take into account
the subtleties associated with the noncompact $SL(2,R)$ group.\\
We have discussed in detail how to construct worldsheet action for compact
coordinates in the case of NSR string on $T^d$. We showed that the equations
of motion can be cast in $O(d,d)$ covariant form since the equations of motion
are conservation laws in the superspace.\\
We  argue that the string coordinates and backgrounds, parametrizing target 
space $AdS_3$ and $S^3$, transform trivially under the T-duality group 
associated with compact directions.Therefore, those equations of motion are
$O(d,d)$ invariant. This completes our study of T-duality symmetry for type 
IIB string on $AdS_3\otimes S^3 \otimes T^4$. 

\bigskip

\noindent {\bf 4. VERTEX OPERATORS FOR EXCITED MASSIVE LEVELS}

\bigskip

\noindent We study construction of duality symmetric vertex operators of NSR 
string in this section. There are interests in excited massive levels of
strings for diverse reasons. It is argued that in Planckian energy scattering
processes stringy states play a very important role in restoring good
high energy behavior of the amplitudes \cite{planck1, planck2}. 
Furthermore, it is conjectured that
there might be enhanced symmetry at these asymptotic energies \cite{gross}. 
There are some 
evidences that excited massive states of closed string might be endowed with
higher symmetries which are not fully explored so far \cite{mv,ov1,ao}. 
Recently, it has been
shown that for bosonic closed string compactified on $T^d$ the vertex operators
associated with excited massive states can be expressed in an $O(d,d)$ 
invariant form. This was achieved in a simple frame work. We worked in the weak
field approximation when strings is considered in the background of massive
excited states. These vertex operators were
first expressed in terms  $\sigma$-derivatives of $X^{\mu}$  and
the  canonical conjugate momenta of the compact coordinates and $\sigma$
and/or $\tau$ derivatives. The vertex operators are required to fulfill 
following conditions. (i) All vertex operators are required
to be $(1,1)$ operators with respect to the stress energy momentum tensors,
$(T_{++}, T_{--})$, of the free string \cite{ov1}. 
This is a powerful constraint
and it leads to the 'equations of motion' and 'gauge conditions'
for the massive backgrounds when they are arbitrary functions of
string coordinates. (ii) At each mass level one constructs
'vertex functions' from the basic building blocks such as 
$\partial X^{\hat\mu},
{\bar \partial}X^{\hat\mu}$ and $\partial ~{\rm or}~ {\bar \partial}$ acting 
on these building blocks. Note that we do not admits terms like 
$\partial{\bar\partial}X^{\hat\mu}~ {\rm or}~ 
{\bar\partial}\partial X^{\hat\mu}$ 
in vertex functions
since these objects and derivative of such objects vanish as a virtue of
free string equations of motion. (iii) The structure of vertex function of a 
given type, at each mass level, is constrained by the level matching conditions
since there is no preferred point on a closed string loop. (iv) The vertex
operator of a given mass level is sum of all such vertex functions. The
vertex operator is required to be $(1,1)$; consequently, at a given mass level,
 the vertex functions are related to each other (see \cite{maha} for details).
At each mass level there are excitations of various 
angular momenta of same mass.
In other words the states belong to the irreducible representations of
$SO({\hat D}-1)$. (v) When we compactify the theory to lower dimensions:
$M_D\otimes T^d$, all the states of a given level are classified according to
irreducible representations of $SO(D-1)$ including the states coming from
excitations in compact directions (these are all scalars) since total degrees
of freedom (at each level) remains the same in both the cases.\\   
In order to construct the desired vertex operators and study their duality
properties , let
us discuss some of the essential ingredients. Consider a closed string in 
background of constant metric
\be
V_G={1\over}\bigg({\hat G}_{\hat\mu\hat\nu}X'^{\hat\mu}X'^{\mu}+
{\hat G}^{\hat\mu\hat\nu}P_{\hat\mu}P_{\hat\nu} \bigg)
\ee
where $P_{\hat\mu}$ is the conjugate momenta. Under the T-duality
$P_{\hat\mu}\leftrightarrow X'^{\hat\mu}$, 
${\hat G}\leftrightarrow {\hat G}^{-1}$. If we included constant ${\hat B}$
background we must have ${\hat M}\leftrightarrow {\hat M}^{-1}$
 under this duality.
In order to see how it works for the first excited massive level consider
the vertex operator for the leading Regge trajectory
\bea 
\label{firstex}
V^{(1)}={\bf {\hat F}}_{{\hat\mu\hat\nu\hat\rho\hat\lambda}}\partial X^{\hat\mu}
\partial X^{\hat\nu}{\bar\partial} X^{\hat\rho}{\bar\partial} X^{\hat\lambda}
\eea
where $ {\bf F}^{(1)}_{{\hat\mu\hat\nu\hat\rho\hat\lambda}} $ is the constant 
background tensor. When we express $\partial X^{\hat\mu}$ and 
${\bar\partial} X^{\hat\mu}$ in terms of 
canonical momenta and $X'$ then (\ref{firstex}) has
 sixteen terms of various types as products of $P$ and $X'$. 
There is a term which is product of
four $P$'s and another term which is product of four $X'$'s as given
below 
\\
\bea
{F^{(1)}}^{{\hat\mu\hat\nu\hat\rho\hat\lambda}}
P_{\hat\mu}P_{\hat\nu}P_{\hat\rho}
P_{\hat\lambda},~~~
F^{(2)}_{{\hat\mu\hat\nu\hat\rho\hat\lambda}}X'^{\hat\mu}X'^{\hat\nu}
X'^{\hat\rho}X'^{\hat\lambda}
\eea
There are eight terms with a set of four terms which is product of 
three $P$'s and one $X'$ and another set of four terms 
with product of three $X'$'s
and a $P$
\be
{F^{(3)}}^{{\hat\mu\hat\nu\hat\rho}}_{\hat\lambda}P_{\hat\mu}P_{\hat\nu}
P_{\hat\rho} X'^{\hat\lambda}, ~~~
{F^{(4)}}^{\hat\lambda}_{{\hat\mu\hat\nu\hat\rho}} X'^{\hat\mu}X'^{\hat\nu}
X'^{\hat\rho} P_{\hat{\lambda}}
\ee
 Finally, there are six terms of the type
where we have product of two $P$'s and two $X'$'s
\be
{F^{(5)}}^{\hat\rho\hat\lambda}_{{\hat\mu\hat\nu}}X'^{\hat\mu}X'^{\hat\nu}
P_{\hat\rho}P_{\hat\lambda}
\ee
Now if we interchange 
$P_{\hat\mu}\leftrightarrow X'^{\hat\mu}$ and look for duality symmetry we
find that $F^{(1)} \leftrightarrow F^{(2)}$, the four tensors of the type
$F^{(3)}$ interchange with precisely the four tensors of $F^{(4)}$ type. 
Finally, the six tensors of $F^{(5)}$ interchange amongst themselves with
appropriate indices \cite{mahaodd}. 
Thus, for constant backgrounds $F^{(i)}$, we have
evidence of the $Z_2$ duality. As a simple example let us work 
in the plane wave
background i.e. the tensors $F^{(i)}$ are multiplied by plane waves. Then 
conformal invariance leads to mass shell conditions and gauge conditions
on these tensors. Moreover, there are three more "vertex functions" for
the first excited massive states in addition to one which corresponds to the 
leading Regge
trajectory. However, when we impose restrictions of conformal invariance only
vertex function associated with leading Regge trajectory survives
for the first excited massive level. Furthermore,
as  we 
consider  higher and higher levels the number of states keep increasing and 
at very  large mass the level degeneracy grows exponentially as is
well known. Therefore,the above procedure to verify conjectured duality becomes
unmanageable. A more efficient technique was introduced \cite{maha}
 to construct duality
invariant vertex operators for closed string. We shall briefly recapitulate
the result and then proceed to generalize the underlying idea for NSR string  
 in what follows. \\
Let us focus our attentions
on vertex functions associated with compact coordinates, $Y^i(\sigma,\tau)$.
(i) It is recognized that the
basic building blocks are $\partial Y^i=P^i+Y'^i,~ P^i$ being the conjugate 
momenta; we have suppressed $\sigma~{\rm and}~ \tau$ dependence of 
$Y^i~{\rm and}~P_i$ here and everywhere from now on. 
Similarly, ${\bar\partial Y^i}=P^i-Y'^i$. These building blocks are
operated upon by $\partial$ and $\bar\partial$ respectively and $\partial\bar
\partial Y^i=0$ by equations of motions. The vertex functions are products
of a string of $P^i\pm Y'^i$ and operators obtained by actions of $\partial,
\bar\partial$. The resulting tensor is contracted with backgrounds of 
appropriate rank tensors which depend only on target space (string) coordinates
$\{X^{\mu} \}$. (ii) The strategy is as follows: recall that $O(d,d)$ vector
is a doublet
\be
V=\pmatrix{ P_i\cr Y'^i\cr }
\ee
where indices are raised and lowered by $\delta^{ij}$ and $\delta_{ij}$. Using
an appropriate projection operator the products of $P^i\pm Y'^i$ 
are re-expressed as products the components
of the $O(d,d)$ vectors $V_i$. Similarly the higher derivatives of these
building blocks can be converted to $O(d,d)$ vectors. The product now
transforms like a tensor under O(d,d) transformations. This tensor can be
contracted with an X-dependent suitable tensor. 
The resulting vertex function is now
$O(d,d)$ invariant. As an example consider a generic vertex function
for the $n^{th}$ massive level
\bea
\label{generic}
{\partial}^pY^{\alpha_i}{\partial^q}Y^{\alpha_j}{\partial^r}
Y^{\alpha_k}...
{\bar\partial}^{p'}Y^{\alpha'_i}{\bar\partial}^{q'}
Y^{\alpha'_j}{{\bar\partial}}^{r'}Y^{\alpha'_k}..,~p+q+r=n+1,~p'+q'+r'=n+1
\eea
The  constraints on $p,q,r, p',q',r'$ follows from the level matching
condition. To convert this tensor to an $O(d,d)$ tensor we refer to
the results of \cite{maha} and quote that 
$\partial^pY=\partial^{p-1}(P+Y'),~~{\bar\partial}^{p'}Y=
{\bar{\partial}}^{p'-1}(P-Y')$. Then using the projection operators, 
$\Delta_{\pm}$, we get \\
$\partial^pY={\Delta_+}^{p-1}\bigg({\bf P}_+V+{\bf\eta}
{\bf P}_-V\bigg),~~{\bar\partial}^{p'}Y={\Delta_-}^{p'-1}
\bigg(({\bf P}_+V-{\bf\eta}{\bf P}_-V\bigg)$, \\
where ${\bf P}_{\pm}$ is
another projection operator and $\eta$ being the $O(d,d)$ metric. Thus
the tensor in (\ref{generic}) can be converted into product of $O(d,d)$ 
vectors. The vertex function is given by
\bea
\label{nlevel}
V_{n+1}={\cal A}_{klm..,k'l'm'..}{\Delta_+}^{p-1}V_+^k
{\Delta}^{q-1}V_+^l
{\Delta_+}^{r-1}V_+^m..{\Delta_-}^{p'-1}V_-^{k'}
{\Delta_-}^{p'-1} V_-^{l'}
{\Delta_-}^{p'-1} V_-^{m'}
\eea
where $V_{\pm}=({\bf P}_+V\pm{\bf\eta}{\bf P}_-V)$ with
$p+q+r=n+1$ and $p'+q'+r'=n+1$.
Note that superscripts $\{k,l,m;k',l',m'\}$ appearing on $V_{\pm}$
in eq. (\ref{nlevel}) are
the indices of the components of the  projected $O(d,d)$ vectors. Moreover,
${\cal A}_{klm..,k'l'm'..}$ is $X$-dependent $O(d,d)$ tensor. Note that
(\ref{nlevel}) will be $O(d,d)$ invariant if coefficients $\cal A$-tensors
transform
as  \\
\bea
\label{rules}
{\cal A}_{klm..,k'l'm'..}\rightarrow \Omega_k^p  \Omega_l^q  \Omega_m^r ...
\Omega_{k'}^{p'} \Omega_{l'}^{q'} \Omega_{m'}^{r'}{\cal A}_{pqr..,p'q'r'..}
\eea
since each term in the product
$\Delta ^{p-1}V^k_+....\Delta^{p'-1}_-V^{k'}$, above, 
transforms like
an $O(d,d)$ vector.\\
We mention in passing that the constraints of conformal invariance need not be
imposed at this stage while we are investigating duality symmetries. Those
requirements further restrict the structures of of vertex operators and
provide useful relations among vertex functions besides imposing mass shell
conditions for a given mass level.
\\
We intend to derive analogous results for the vertex functions of the excited
massive states of NSR string. Notice that for the first excited level on
the leading Regge trajectory for NSR string will have a lot 
more terms compared
to (\ref{firstex}) since we can construct additional terms which contract
with chiral worldsheet fermions. For example, we can have generic terms like
\bea
\label{superf} 
G^{(1)}_{\hat\mu\hat\nu\hat\rho\hat\lambda\hat\delta}\partial X^{\hat\mu}
\partial X^{\hat\nu}\psi^{\rho}_-\psi^{\hat\lambda}_-
{\bar\partial}X^{\hat\delta}
,~ G^{(2)}_{\hat\mu\hat\nu\hat\rho\hat\lambda\hat\delta\hat\epsilon}
\partial X^{\hat\mu}\partial X^{\hat\nu}\psi^{\rho}_-\psi^{\hat\lambda}_-
\psi^{\delta}_-\psi^{\hat\epsilon}_-
\eea
and several other terms where $\partial$ is replaced by ${\bar\partial}$ and
$\psi_-$ is replaced by $\psi_+$ so long as we ensure, to start with, we have
maintained same dimensionality for product of left movers and right movers with
respect to the two stress energy momentum tensors. The discrete $Z_2$ symmetry
alluded
to in (\ref{z2duality}) will be maintained if we take into account all required
terms for the vertex operator under considerations. It is quite obvious even
keeping track of all terms for the vertex operators of some of the low lying
excited
massive levels is going to be not very efficient if we want to check
the conjectured T-duality for superstrings in terms of the bosonic coordinates
and NSR fermions. So far there is no construction of manifestly $O(d,d)$
invariant vertex operators for NSR string along compact directions even for 
the massless sector i.e. massless scalars that arise from compactification
of $\hat G$ and $\hat B$.\\
Therefore, we resort to the superfield approach and consider vertex functions
for massive excited states constructed out of the superderivatives of
superfields. There are two types of generic vertex functions
\\
(i) ${\overline D}W^{i_1}{\overline D}W^{i_2}... {\overline D}W^{i_m} 
DW^{j_1}DW^{j_2}...
 DW^{j_m}$.These 
correspond to leading Regge trajectories.\\
(ii) ${\overline D}^pW^{i_1} {\overline D}^qW^{i_2}....{\overline D}^r
W^{i_m}{D}^{p'}W^{j_1}
{D}^{q'}W^{j_2}...{D}^{r'}W^{j_m}$ and we require $p+q+r=p'+q'+r'$. \\
We proceed with following step for the case (i) 
(I) Recall that $\bf U$ is an $O(d,d)$ vector, whose upper component is
 $W^i$ and
the lower component is its dual ${\widetilde W}_i$. Introduce two projection
operators
\bea
\label{uproject}
{\widetilde P}_+=\pmatrix{1 & 0\cr 0 & 0\cr} ~~~
{\widetilde P}_-=\pmatrix{0 & 0\cr 0
& 1 \cr}
\eea
Note that ${\widetilde P}_+{\bf U}=W$. \\
(II) Introduce a doublet through the pair 
$(D,{\overline D})$.
\be
\label{ddoublet}
 {\cal D}=\pmatrix{D \cr {\overline D} }
\ee
Then  projection operators
\bea
\label{projectD}
{\widetilde \Delta}_+=\pmatrix{1 & 0 \cr 0 & 0 \cr} ~~{\widetilde \Delta}_-=
\pmatrix{0 & 0 \cr 0 & 1 \cr}
\eea
Note that ${\widetilde P}_{\pm}$ are $2d\times 2d$ 
dimensional projectors whereas
${\widetilde \Delta}_{\pm}$ are $2\times 2$ projectors.\\
The vertex functions which assume the form given in (i) above can be cast as
products of $O(d,d)$ vectors
\bea
\label{oddtensor1}
{\widetilde \Delta}_-{\cal D}{\widetilde P}_+{\bf U}^{\alpha_1}....
{\widetilde \Delta}_-{\cal D}{\widetilde P}_+{\bf U}^{\alpha_m} 
{\widetilde \Delta}_+{\cal D}{\widetilde P}_+{\bf U}^{\beta_1}....
{\widetilde \Delta}_+{\cal D}{\widetilde P}_+{\bf U}^{\beta_m}
\eea
Thus we have an $O(d,d)$ tensor of rank $2m$. We contract it with
a tensor, which depends on spacetime superfield $\phi^{\mu}$ to construct an
$O(d,d)$ invariant vertex function for the $n^{th}$ massive level.
\bea
\label{leadingv}
V_{n+1}=T_{\alpha_1..\alpha_m\beta_1..\beta_m}
{\widetilde \Delta}_-{\cal D}{\widetilde P}_+{\bf U}^{\alpha_1}....
{\widetilde \Delta}_-{\cal D}{\widetilde P}_+{\bf U}^{\alpha_m}
{\widetilde \Delta}_+{\cal D}{\widetilde P}_+{\bf U}^{\beta_1}....
{\widetilde \Delta}_+{\cal D}{\widetilde P}_+{\bf U}^{\beta_m}
\eea
If the $O(d,d)$ vector transforms as: 
$U^{\alpha_1}\rightarrow \Omega^{\alpha_1}_{\alpha '_1}U^{\alpha'_1}$, then we
require
\bea
\label{trans}
T_{\alpha_1,..\alpha_m\beta_1..\beta_m}\rightarrow \Omega^{\alpha'_1}_{\alpha_1}
..\Omega^{\alpha'_m}_{\alpha_m}\Omega^{\beta'_1}_{\beta_1}...
\Omega^{\beta'_m}_{\beta_m}T_{\alpha'_1..\alpha'_m\beta'_1...\beta'_m}
\eea
so that the vertex function $V_{n+1}$ is T-duality invariant.\\
Now we focus attention on the second type of vertex function mentioned in (ii)
above. Note that a typical term appearing in the product is like
$({\overline D})^p$ and $(D)^{p'}$. 
We can use the projection operators introduced
in (I) and (II) above to express products of such terms as
\bea
\label{odds2}
{\widetilde\Delta}_-^p{\widetilde P}_+U^{\alpha_1}{\widetilde\Delta}_-^q
{\widetilde P}_+
U^{\alpha_2}..{\widetilde\Delta}_-^r{\widetilde P}_+U^{\alpha_m}
{\widetilde\Delta }_+^{p'}{\widetilde P}_+U^{\beta_1} 
{\widetilde\Delta }_+^{p'}{\widetilde P}_+U^{\beta_2}.. 
{\widetilde\Delta }_+^{p'}{\widetilde P}_+U^{\beta_m} 
\eea
Now this is an $O(d,d)$ tensor of rank $p+q+r$ satisfying the level matching 
condition. As in the previous case, we have to just contract with a tensor
(which depends on superfield $\phi$) to get an $O(d,d)$ invariant vertex
function.\\
So far we have left out two other possibilities which we dwell upon now.
There are two more types of vertex functions in a given level: \\
(a) We can have a situation that the vertex function has product of mixed
set of operators i.e. some of the superfields correspond to spacetime
coordinates and some to compact ones.
\bea
\label{mixedvertex}
T_{\mu_1..\mu_k\alpha_1..\alpha_l\mu'_1..\mu'_k\alpha'_1\alpha'_l}
{\overline D}^p\phi^{\mu_1}..{\overline D}^q\phi^{\mu_k}{\overline D}^{r}
W^{\alpha_1}{\overline D}^s
W^{\alpha_l}D^{p'}\phi^{\mu'_1}D^{q'}\phi^{\mu'_k}D^{r'}W^{\alpha'_1}D^{s'}
W^{\alpha'_l}
\eea
First notice that $\phi^{\mu}$ is inert under T-duality transformations. 
Similarly, all the spacetime indices of the tensor $T_{\mu\nu...\mu'\nu'..}$ 
do not get transformed under T-duality. Moreover, all the spacetime indices
of $T$ are contracted with spacetime superfields so that effectively we deal
with a tensor with "internal" indices which are contracted with product of
building blocks consisting of superderivatives of $W$'s i.e ${\overline D}~ 
{\rm or}~ D$
acting on $W$'s . We already presented a prescription of constructing $O(d,d)$
invariant vertex functions out of such products. Therefore, any arbitrary
vertex function of a given massive level can be expressed in an $O(d,d)$
invariant form.
\\
(b) There is another class of vertex functions which are product of the
superderivatives of the spacetime superfields only. However, this class of
vertex functions are automatically T-duality invariant since the 
superderivatives $\phi^{\mu}$ and corresponding $\phi$-dependent tensors are
not sensitive to $O(d,d)$ transformations.\\
In conclusion for an NSR string compactified  on $T^d$, we can express all
vertex functions at each massive level in T-duality invariant form.
\\
We address another point in the context of type IIB theory.
It is well known that this theory
 in endowed with S-duality symmetry. Its massless
spectrum in NS-NS sector for 
critical dimension (${\hat D}=10$) consists of graviton, 
${\hat g}_{\hat\mu\hat\nu}$, 2-form antisymmetric tensor, 
${\hat B}^{(1)}_{\hat\mu\hat\nu}$ and dilaton, ${\hat{\cal\phi}}$. The R-R
sector is axion, ${\hat{\cal \chi}}$, 2-form antisymmetric tensor, 
${\hat B}^{(2)}_
{\hat\mu\hat\nu}$ and a four form tensor 
${\hat C}^{(4)}_{\hat\mu\hat\nu\hat\rho\hat\lambda}$ whose field strength
is required to be self dual. The effective action for the type IIB theory
may be expressed in an S-duality invariant form. When we toroidally
compactify the theory to lower dimension, the reduced effective can also
written in S-duality invariant form. Therefore, starting from NS-NS 
backgrounds, we can generate RR backgrounds of the reduced theory; however,
the reduced tensors of four form $C^{(4)}$ cannot be generated from
the NS-NS backgrounds. Let us closely examine  
 the case 
when type IIB theory is compactified on $T^4$ to a six dimensional 
theory and focus our attention on the moduli and the vector fields
coming from reduction of backgrounds
\footnote{ I am thankful to John Schwarz for elucidating the arguments
presented here} .
 We expect that the
massless states coming from NS-NS sector will be classified according to
representations of $O(4,4)$. In fact the moduli parametrizes the coset
${{O(4,4)}\over{O(4)\otimes O(4)}}$ as  was demonstrated by us \cite{ms}.
The counting is quite simple: the moduli coming from compactification of the
graviton and the 2-form antisymmetric tensor  add up to 16 as
expected. The gauge fields originating from metric and antisymmetric tensor
(from NS-NS sector) belong to the vector representation of $O(4,4)$. Indeed,
in the NS-NS sector the worldsheet action exhibits presence of all these
massless fields, if we follow prescriptions of ref \cite{ms}. Let us turn to
the R-R sector. There are 9 scalar appearing due to compactification of the 
2-form, $B^{(2)}$, the 4-form $C^{(4)}$ besides the axion. The number of vector
fields are eight: four from $B^{(2)}$ and four from $C^{(4)}$. We should also
take into account the underlying S-duality symmetry: ${SL(2,Z)}$.
It is more appropriate to classify massless states of the toroidally 
compactified six dimensional theory combining the states
from NS-NS and RR sector. The
arguments are along the same line as classification of branes (hence 
classifying the background tensors) in the
context of toroidal compactification of type IIB theory and 
M-theory \cite{jhs1,jhs2,jm2}.
 They belong to representations of $O(5,5)$ from this
perspective. Whereas the 25 (=16+9) moduli parametrize the coset
${O(5,5)}\over{O(5)\otimes O(5)}$, the 16 (=8+8) massless vectors belong 
to the spinor representation of $O(5,5)$.The other backgrounds, in the
six dimensional theory,  also belong
to appropriate representation of this group.\\
 We note that one can study the T-duality attributes of the NS-NS massless 
backgrounds of the theory compactified on $T^4$ in the worldsheet approach
presented here. The massive excited backgrounds along the compact direction,
in the NS-NS sector, can be coupled to corresponding worldsheet 
supercoordinates. We are able to express the vertex operators for each of 
such levels in a manifestly  T-duality invariant form. However, it is not
possible to construct similar vertex operators for the RR sector in the 
present formulation.

\bigskip

\noindent{\bf 5. SUMMARY AND CONCLUSIONS}

\bigskip

\noindent In this article we focused our attention to investigate T-duality
properties of NSR string in its massless backgrounds. We have shown that the
worldsheet equation of motion of the NSR string compactified on $T^d$ can 
be expressed in an $O(d,d)$ covariant form in the presence of backgrounds. 
In order to achieve this objective
it is most appropriate to adopt the worldsheet superfield formalism. The
equations of motion corresponding to superfields along compact directions
were recognized to be conservation laws when the background components along
compact directions depend only on spacetime superfield coordinates. We 
introduced a set of dual superfields and a set of dual backgrounds to write
a corresponding dual action. The equation of motion for this case is also 
expressed as a conservation law. It is shown that the two sets of equations 
of motion can be judiciously combines and cast in an $O(d,d)$ covariant form.
\\
We generalized an earlier  prescription for closed bosonic string
case to construct vertex operators for massive excited levels of NSR string
in the superspace formulation.
It was shown that the vertex operators for each massive level can be cast
in an $O(d,d)$ invariant form. Moreover, for a special case when type IIB
theory is compactified on $AdS_3\otimes S^3\otimes T^4$, it is possible to
construct vertex operators for excited massive levels in $O(4,4)$
invariant form. This is an improvement over our result for the closed bosonic
string case where the spacetime geometry was taken to be flat.

\bigskip
\noindent
{\bf Acknowledgments:} It is a great pleasure to acknowledge many valuable
discussions with John Schwarz. I thank him for his keen interests in this work
 and for sharing his deep insights and for carefully reading the 
manuscript and bringing references on double field theory to my attention. 
I would like to thank John Schwarz and the Theory Group
at Caltech for their very gracious and  warm hospitality during the
winter term of 
2011. My visit to Caltech was supported by Department of Energy Grant
DE-FG02-92ER40701. This work is generously supported by the People of
the Republic of India and through a Raja Rammanna Fellowship of DAE, India.

\end{document}